\documentclass[review]{elsarticle}

\journal{Optics Communications}

\usepackage{graphicx}
\usepackage{subfigure}
\usepackage{caption}
\biboptions{numbers,sort&compress}
\usepackage{upgreek}
\usepackage{amsmath}

\begin{document}
	
	\begin{frontmatter}
		
		\title{Launch and capture of a single particle in a pulse-laser-assisted dual-beam fiber-optic trap}

		
		\author{Zhenhai Fu}
		\author{Xuan She}
		\author{Nan Li}
		\author{Huizhu Hu\corref{cor1}}
		\ead{huhuizhu2000@zju.edu.cn}
		\cortext[cor1]{Corresponding author}
		
		\address{State Key Laboratory of Modern Optical Instrumentation, College of Optical Science and Engineering, Zhejiang University, Hangzhou, 310027 China}

		\begin{abstract}
			The rapid loading and manipulation of microspheres in optical trap is important for its applications in optomechanics and precision force sensing. We investigate the microsphere behavior under coaction of a dual-beam fiber-optic trap and a pulse laser beam, which reveals a launched microsphere can be effectively captured in a spatial region. A suitable order of pulse duration for launch is derived according to the calculated detachment energy threshold of pulse laser. Furthermore, we illustrate the effect of structural parameters on the launching process, including the spot size of pulse laser, the vertical displacement of beam waist and the initial position of microsphere. Our result will be instructive in the optimal design of the pulse-laser-assisted optical tweezers for controllable loading mechanism of optical trap.
		\end{abstract}
		
		\begin{keyword}
			{optical trap}\sep {pulse laser}\sep {particle manipulation}
		\end{keyword}
		
	\end{frontmatter}
	
	
	\section{Introduction}
	Since the first demonstration of optical trapping and manipulation of dielectric particles by Ashkin and his co-workers in 1970s \cite{ashkin1970acceleration,ashkin1971optical,ashkin1976optical}, optical manipulation has been developed rapidly for nearly 50 years,  ranging from biology (e.g.,\cite{ashkin1987optical,ashkin1987optical2,neuman2004optical}) to fundamental physics \cite{geraci2010short,moore2014search,rider2016search}. Optically levitated and cooled dielectric micro-spheres in high vacuum show great potential as resonant force detectors 
	\cite{ranjit2015attonewton,yin2013optomechanics,geraci2010short,ranjit2016zeptonewton} and light force accelerometer (LFA) as proposed previously 
	\cite{butts2008development,kotru2010toward,yu2010theory,li2015simulation}. To achieve engineering applications, a key technology is to realize efficient and controllable loading process, especially repeatable launch of sensing particle. However, the traditional two methods with shaking mechanism \cite{butts2008development,li2010measurement,arita2013laser,moore2014search,ranjit2015attonewton} or nebulizer source \cite{kiesel2013cavity,summers2008trapping,grass2013optical,gieseler2014dynamics} are essentially random. They are unable to decide which particle to be captured or how many particles are trapped. To maintain capture efficiency, one should spray excessive samples into the optical trap region, while the residual particles will inevitably become contamination of vacuum and optics.

	In contrast, the optical loading methods are more controllable. Pulsed-laser has been widely used in the field of laser cleaning, which produces a strong lifting force that could overcome adhesion forces and gravity to remove the particles quickly and efficiently \cite{lu1996laser,lu1994surface,kane2006laser} without considering the laser-induced damage to the particles. Moreover, pulse-laser-assisted optical tweezers (PLAT) open the possibility of manipulating stuck microparticles in aqueous environments and may find broad applications in cell biology and molecular biology
	\cite{ambardekar2005optical,sugiura2012pulse,maeda2012optical}. Rapid loading and manipulation of a microsphere in air or vacuum remains challenge until now. The dual-beam optical trap based on micro-structure offers intrinsic benefits of miniaturization and integration, which is crucial for developing a practical sensor
	\cite{jensen2005demonstration,barbour2010inertial,black2012fiber,bellini2010femtosecond,su2016chip,li2016dynamic}. 
	
	In this work, we propose a new structure of PLAT that combinates pulse laser with dual-beam fiber-optic trap. We develop a three-dimensional numerical model for the launch and capture process of particles in a dual-beam fiber-optic trap with assistance of pulse laser. The radiation forces generated by CW beams and pulse beam are described based on ray optics approximation. Taking the gravity and viscous force of air into account, we first point out the effective capture region of dual-fiber optical trap. Furthermore, the effect of pulse duration is analyzed from a quantitative perspective of energy transfer and a suitable order of pulse duration for launch in air condition is derived according to the calculated energy threshold. Finally, the influence of other parameters of optical structure on detachment threshold is revealed in a more complex 3D situation, including the spot size, the vertical displacement of beam waist and the initial location of microspheres. 
	
	\section{Pulse-laser-assisted dual-fiber optical trap}
	\subsection{Modelling geometry and materials}
\begin{figure}
		\centering
		\includegraphics[scale=0.5]{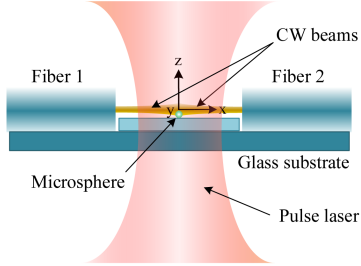}
		\caption{Schematics of pulse laser assisted dual-fiber optical trap}
		\label{Schematics}
	\end{figure}

	Figure \ref{Schematics} shows the schematics of a pulse laser assisted dual-fiber optical trap. The target particle initially adheres to the glass substrate within the effective capture region of the optical trap. The particle cannot be levitated simply by the gradient force of two CW beams before launched by the focused pulse beam because of the adhesion forces \cite{van1989colloidal}.
	The propagation direction of pulse beam is vertically upward and its waist is at the same position as the plane where the particle lies. During the irradiation of pulse beam, the particle not only obtains a vertically upward velocity from the scattering force, but also be pushed towards the center of the beam spot by the gradient force. To be free from the substrate, the launched particle must move a sufficient distance away from the surface compared with the effective limits of the adhesion forces. 
	
	The CW trapping beams are emitted from different lasers to avoid the generation of coherent interference. The power from each fiber end is $P_1=P_2=200\,\rm{mW}$ with the spacing of $200\,{\rm{\upmu m}}$, and they are both regarded as Gaussian mode. We assume that the beam waist is located at the fiber end and the waist radius equals to the core radius $\omega_0=2.5\,\rm{\upmu m}$. The wavelengths of the trapping laser and pulse laser are $\lambda_0=980\,\rm{nm}$  and $\lambda_p=1064\,\rm{nm}$ respectively. In the study, a polystyrene microsphere (refractive index $n_0=1.59$, radius $a=5\,\rm{\upmu m}$, density $\rho=1.05\times10^3\,\rm{kg/m^3}$) is selected as the representative target. And its viscosity coefficient is $\eta=1.8\times10^{-5}\,\rm{Pa\cdot s}$ under normal temperature and pressure conditions in air.
	
	\subsection{Mechanical model}
	As shown in figure \ref{Schematics}, the origin of coordinate system is set at the symmetry center of optical trap. Two fiber-emitted beams counter propagate along horizontal $x$ direction, while the pulse beam propagates along vertical $z$ direction. The dynamic trajectory of the launch and capture process microsphere can be modeled by the Newton's equation:
	\begin{equation}
	m\mathbf{\ddot{r}}=\mathbf{F_{cw}(r)}+\mathbf{F_{pulse}(r},t)+\mathbf{F_{S}}(t)+\mathbf{F_{A}(r)}+m\mathbf{g} \label{eq1}
	\end{equation}
	where $\mathbf{r}$, $m$ represent the position and mass of the microsphere respectively. The radiation forces generated by CW beams $\mathbf{F_{cw}(r)}$ and the adhesion forces $\mathbf{F_{A}(r)}$ are only related to the position of microsphere in optical fields, while the radiation forces generated by pulse beam $\mathbf{F_{pulse}(r},t)$ and the viscous force $\mathbf{F_{S}}(t)$ vary with time. Unlike in water, where the buoyancy cancels out gravity, here, we take the gravity force of microsphere into account. Although the gravity is much smaller than the radiation forces when the particle is near the optical trap, particle will fall and return to the trapping area under the action of gravity once it has been launched much higher above the optical trap. We also ignore the random force due to Brownian motion and the lifting force caused by rapid thermal expansion of the substrate in the simulation because the extinction coefficients and thermal conductivities of both particle and glass substrate are relative small \cite{chase1998nist}.
	
	Van der Waals force is typically the predominant adhesion force for micron-size particles \cite{van1989colloidal}, including forces between molecules possessing dipoles and quadrupoles caused by the polarizations of the atoms and molecules in the material. Here, we neglect the deformation of particle caused by the Van der Waals force. A capillary force will be negligible while calculating the adhesion force of dry particles because there is no liquid layer between the particle and substrate. Therefore, the magnitude of total adhesion forces can be approximately expressed as \cite{mittal1988particles}
	\begin{equation}
	F_A=\frac{Aa}{6Z^2}
	\label{eqFA}
	\end{equation}
	
	The Hamaker constant $A$ is typically in the order of $10^{-21}\sim10^{-19}\,\rm{J}$, and we set it to $10^{-19}\,\rm{J}$ in the simulation for simplicity. The separation between the particle and surface $Z$ varies as particle is launched by pulse laser, and its initial value is the atomic separation of approximately 4 angstroms \cite{mittal1988particles}. 
	The viscose force due to the viscosity of air is proportional to the instantaneous velocity of the particle $\mathbf{v}(t)$:
	\begin{equation}
	\mathbf{F}_S(t)=-6\pi a\eta\mathbf{v}(t)
	\end{equation}
	
	The size of the target microsphere is much greater than the wavelengths of lasers, hence the optical forces on the microsphere can be calculated by ray optics \cite{ashkin1992forces,li2016theoretical}. The optical force acting on the sphere is the sum of contribution of each single ray, which can be divided into the scattering force component $d\mathbf{F}_s$ and the gradient force component $d\mathbf{F}_g$:
	\begin{equation}
	\left\{
	\begin{aligned}
	d\mathbf{F}_s&=\frac{n_1q_s}{c}\mathbf{s}dP \\       d\mathbf{F}_g&=\frac{n_1q_g}{c}\mathbf{g}dP
	\end{aligned}
	\right.
	\end{equation}
	where $n_1$ is the refractive index of the air ($n_1=1$ in the simulation), $c$ the speed of light in vacuum, $dP$  the power of single rays. $q_s$ and $q_g$ are the efficiency factors of each part, which are given by:
	\begin{equation}
	\left\{
	\begin{aligned}
	q_s&=1+R\cos2\alpha_i-T^2\frac{\cos(2\alpha_i-2\alpha_r)+R\cos2\alpha_i}{1+R^2+2R\cos2\alpha_r}       \\ 
	q_g&=-1+R\sin2\alpha_i+T^2\frac{\sin(2\alpha_i-2\alpha_r)+R\sin2\alpha_i}{1+R^2+2R\cos2\alpha_r} 
	\end{aligned}
	\right.
	\end{equation}
	
	where $\alpha_i$ and $\alpha_r$ are the angles of incidence and refraction. $R$ and $T$  are the reflection and the transmission coefficients at the surface of the microsphere. The beam intensity of two fiber-emitted beams and the focused pulse beam can be described with Gaussian expression:
	\begin{equation}
	\left\{
	\begin{aligned}
	I_{CW}(x,y,z)&=I_0\frac{\omega_0^2}{\omega_0(z)^2}\exp[\frac{-2(x^2+y^2)}{\omega_0(z)^2}]          \\
	I_{pulse}(x,y,z,t)&=I_{pulse}(t)\frac{\omega_p^2}{\omega_p(z)^2}\exp[\frac{-2(x^2+y^2)}{\omega_p(z)^2}] \\
	\end{aligned}
	\right.
	\end{equation}
	\begin{equation}
	\left\{
	\begin{aligned}
	\omega_0(z)=\omega_0\sqrt{1+(\frac{\lambda_0z}{\pi\omega_0^2})^2} \\
	\omega_p(z)=\omega_p\sqrt{1+(\frac{\lambda_pz}{\pi\omega_p^2})^2}
	\end{aligned}
	\right.
	\end{equation}
	
	The intensity of CW beams $I_0$  is constant and a typically temporal Gaussian pulse shape for pulse laser was used in the study \cite{bauerle2000analysis}：
	\begin{equation}
	I_{pulse}(t)=I_p(\frac{t}{\tau_0})^\beta\exp[\beta(1-\frac{t}{\tau_0})]
	\end{equation}
	
	where $I_p$ and $\tau_0$ is the peak intensity and the pulse duration, respectively. The temporal shape factor is defined as $\beta=1$.
	We first calculate the value of each force in Eq.(\ref{eq1})  under the same coordinate, and then decompose the forces into the three directions of the coordinate system:
	\begin{equation}
	\left\{
	\begin{aligned}
	m\mathbf{\ddot{z}}&=\mathbf{F_{cw-z}(z)}+\mathbf{F_{pulse}(z},t)+\mathbf{F_{S-z}}(t)+\mathbf{F_{A}(z)}+m\mathbf{g} \\
	m\mathbf{\ddot{x}}&=\mathbf{F_{cw-x}(x)}+\mathbf{F_{pulse}(x},t)+\mathbf{F_{S-x}}(t) \\
	m\mathbf{\ddot{y}}&=\mathbf{F_{cw-y}(y)}+\mathbf{F_{pulse}(y},t)+\mathbf{F_{S-y}}(t) \\
	\end{aligned}
	\right.
	\label{eq3D}
	\end{equation}	
	
	where the $x$, $y$ and $z$ subscripts in each term represent force component in three coordinate axes. We assume that the adhesion force only contains component in the z-axis direction. Based on the Simulink toolbox, we simulate the dynamic behavior of microsphere under different initial conditions.
	
	\section{Results and discussion}
	\subsection{Effective region of dual-fiber optical trap}
	To select the appropriate initial positions of the microspheres in the optical field, the effective region of the dual-fiber optical trap is simulated first. Here, the effective region can be characterized as an area of optical field where any stationary particle can be lead to the equilibrium point under the action of optical force and gravity. Apparently, a launched microsphere will fall back to the substrate if it doesn't enter this area. And microspheres with high initial speed will escape from the trap even though getting into this area. Therefore, only microspheres entering and hovering in this area can be trapped by the optical trap.
	
\begin{figure}
		\centering
		\subfigure[]{
			\label{FIG2(a)} 
			\includegraphics[width=0.95\textwidth]{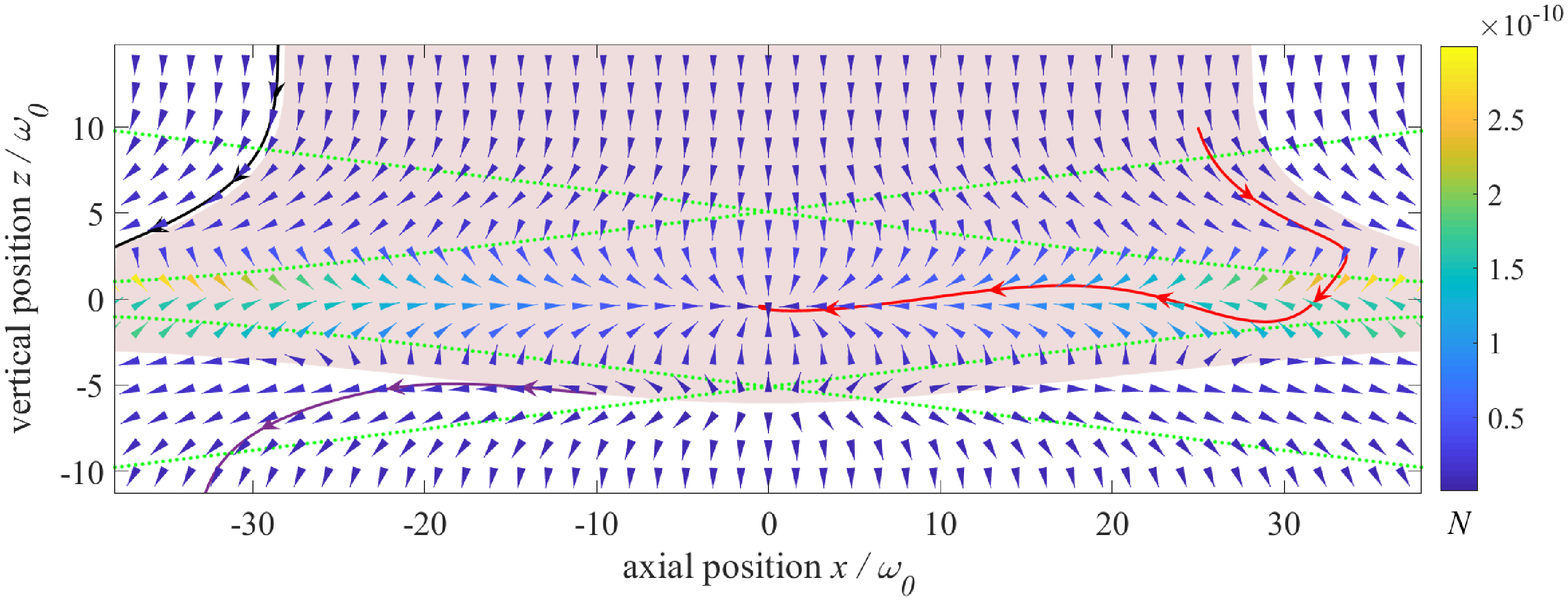}}
		\subfigure[]{
			\label{FIG2(b)}
			\includegraphics[width=0.95\textwidth]{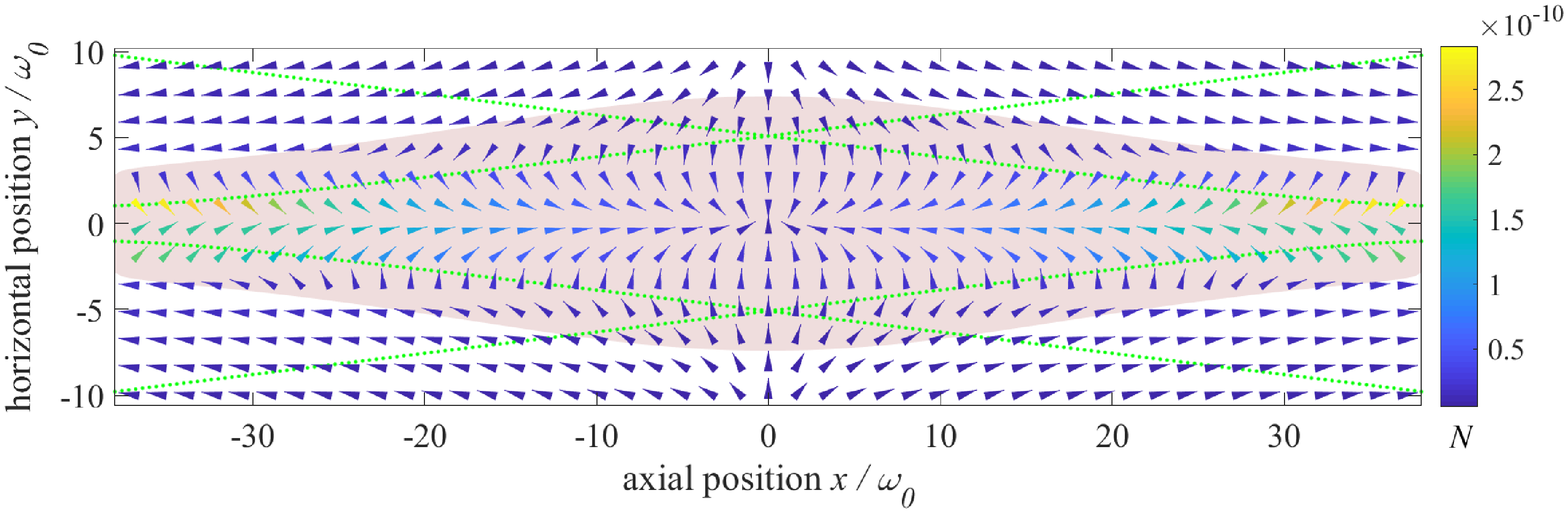}}
		\subfigure[]{
			\label{FIG2(c)}
			\includegraphics[width=0.95\textwidth]{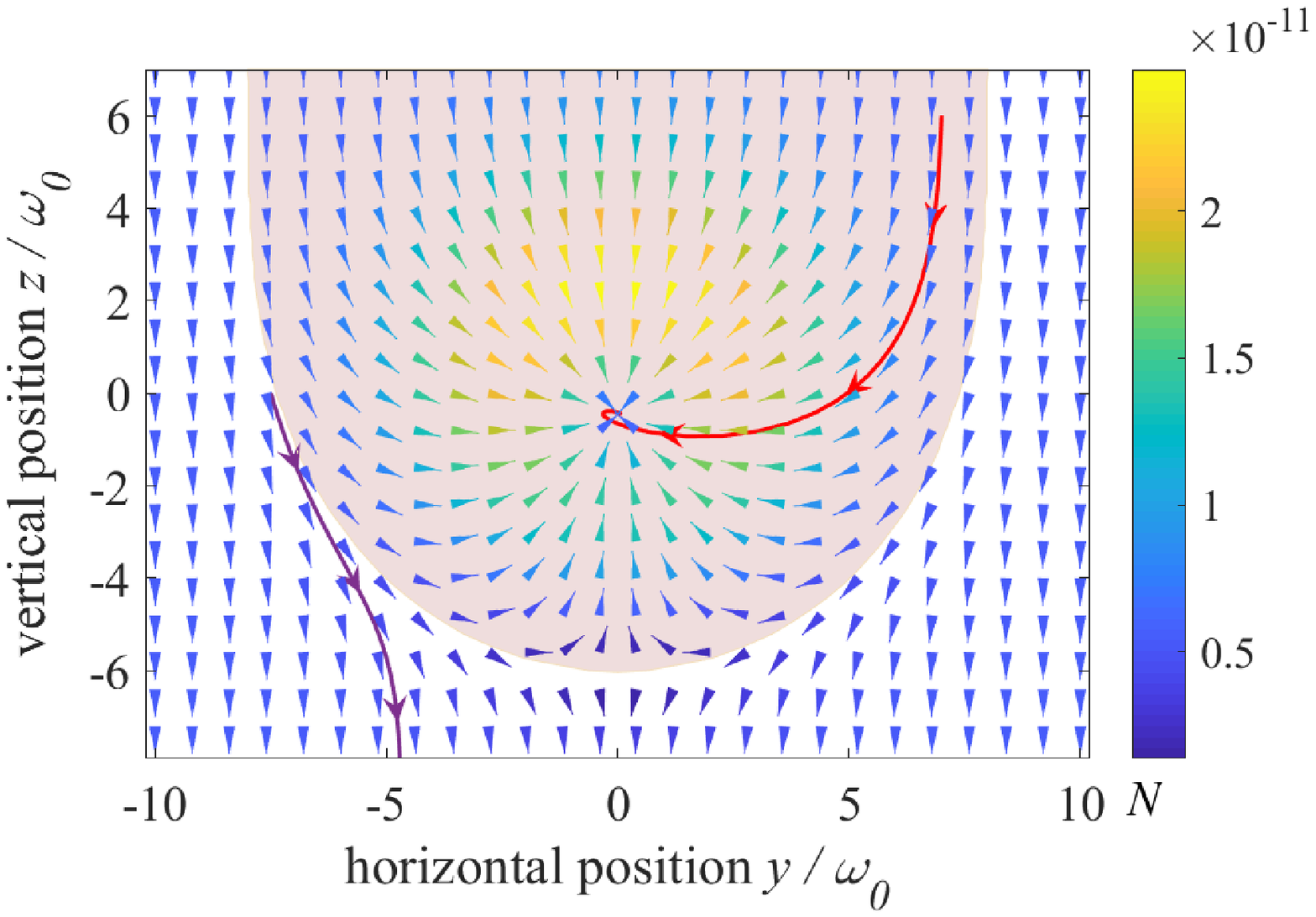}} 
		\caption{Simulated summation of optical force and gravity exerted on the microsphere in three sections: (a) the vertical section, (b) the horizontal section and (c) the radial section. Pink shaded area: the effective region in each section. Red solid curves: capturing trajectory of a microsphere in the effective region. Black solid curve: the trajectory of a microsphere hitting the fiber end outside the effective region. Purple solid curves: the escaping trajectory of microspheres beneath the effective region. Green dotted curves: the beam width from two fibers. The directions and colors of the arrows in (a) or (c) respectively represent the directions and magnitudes of the resultant forces. The perpendicular components of resultant forces in (b) are non-zero because of the gravity, and the directions of the arrows merely represent the directions of the force components in this plane while the colors represent the magnitudes of the entire forces.}
		\label{FIG2}		
	\end{figure}

	Figure \ref{FIG2} show the simulated summation of optical force and gravity exerted on the microsphere in three planes: the vertical section of $y=0$ , the horizontal section of $z=0$  and the radial section of  $x=0$. The directions and colors of the arrows represent the directions and magnitudes of the resultant forces, respectively. According to Eq.(\ref{eq3D}) and eliminating terms of pulse beam and adhesion forces, we simulated the dynamic behavior of the microsphere of different initial positions in the space. And the effective region within each plane can be drawn from the simulated results, as shown in the pink shaded area. 
	
	Due to vertical downward gravity, the lower boundary and the upper boundary of the effective region in vertical section are asymmetry. Below the lower boundary, the gradient component of optical forces is too weak to offset gravity, and microspheres escape from the trap as shown in the purple solid curve. In the center of the symmetry ($x=0$), where the optical forces get its maximum coverage, the maximum vertical distance between the edge and the beam axis is $\Delta z=6.02\omega_0=15.05\,\upmu\rm{m}$. This is the lowest point in space where microspheres can be captured. When the difference of beam widths increases with $|x|$, the effective range narrows down. The vertical distance gets minimum value of $\Delta z=2.82\omega_0=7.05\,\upmu\rm{m}$ when microspheres are close to fiber ends. On the contrary, for the upper boundary, once $|x|\le27.6\omega_0=69\,\upmu\rm{m}$, microspheres can always get into the trap under the action of gravity even beyond the upper boundary of the fibers. The velocity of the microsphere will reach a maximum value during the descent process, and they can still be effectively captured at such an equilibrium speed when they enter the optical field.
	
	By contrast, the resultant forces appear as symmetrical distributions in horizontal section and radial section. The effective region in horizontal section is spindle shaped, and the closer to the fiber end, the narrower it behaves, as indicated in Fig.\ref{FIG2(b)}. A maximum horizontal distance of $\Delta y=8.13\omega_0=20.3\,\upmu \rm{m}$ is about twice the diameter of microspheres as shown in Fig.\ref{FIG2(c)}, while the minimum horizontal distance is about $\Delta y=2.82\omega_0=7.06\,\upmu\rm{m}$, slightly bigger than the microsphere radius.
	
	The irregular distribution of effective region can be largely ascribed to the spatial nonuniformity of two CW beams. The widths of two beams are equal only in the equilibrium plane of $x=0$, where the scattering forces at different radial distances cancel each other and the summations of gradient forces point to the equilibrium center. The difference of two beam widths increases as it is close to fiber end, resulting in different condition of scattering forces at different radial distances. When it is close to the beam axis (i.e. radial distance of $|z|$ or $|y|$ is small), the scattering force of the narrower beam is predominant thus their summations point to the equilibrium center and microspheres can reach the equilibrium center under the action of the optical trap. Instead, when the radial distance is much larger than the narrower beam width, the scattering forces of the larger one is dominant over another. The resultant forces point to fiber ends, and the microspheres will hit the end finally. The black solid curve in Fig.\ref{FIG2(a)} denotes the trajectory of microsphere for this case.
	
	\begin{figure}
		\centering
		\subfigure[]{\label{FIG3(a)}\includegraphics[width=\textwidth]{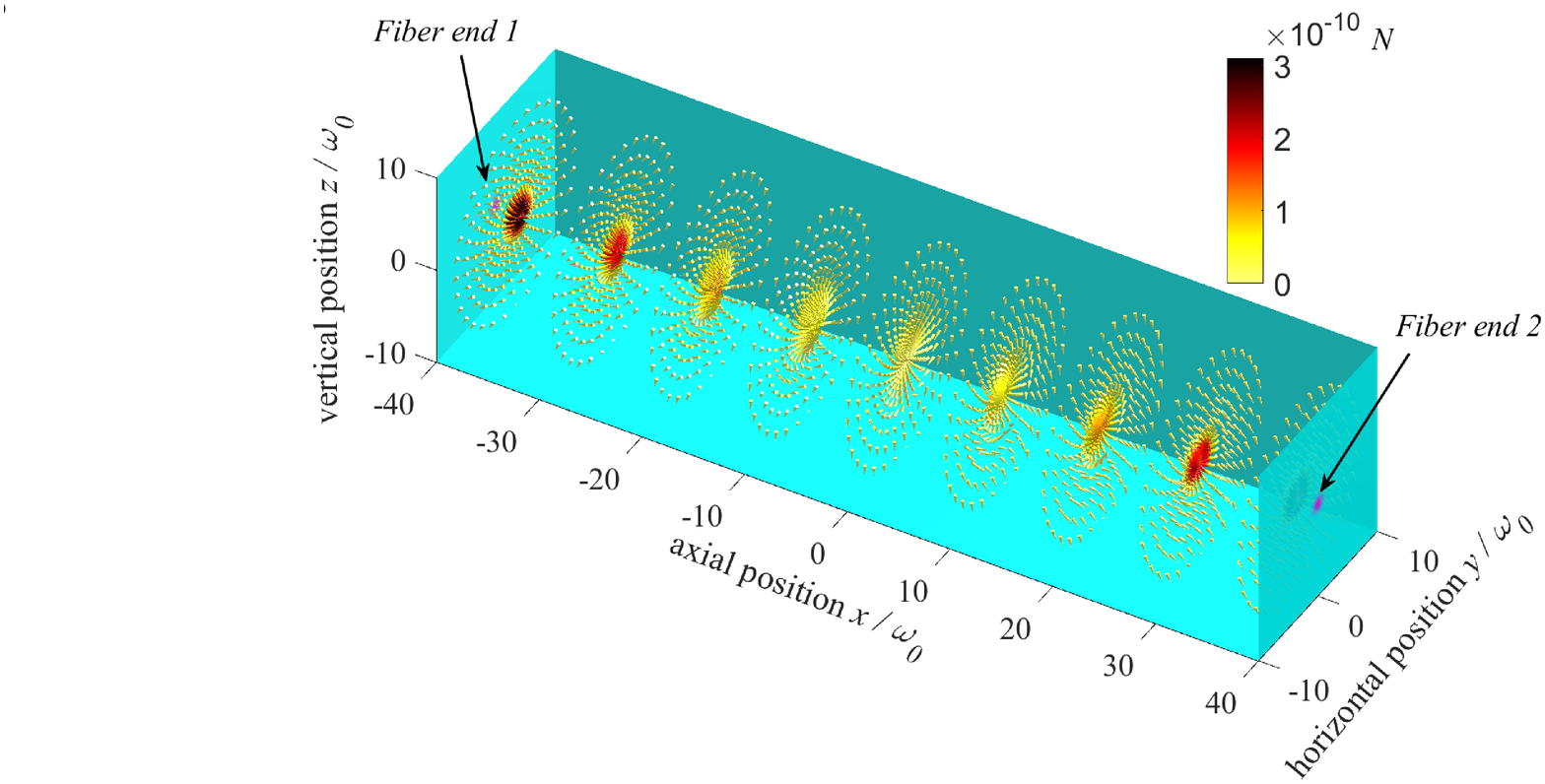}}
		\subfigure[]{\label{FIG3(b)}\includegraphics[width=0.53\textwidth]{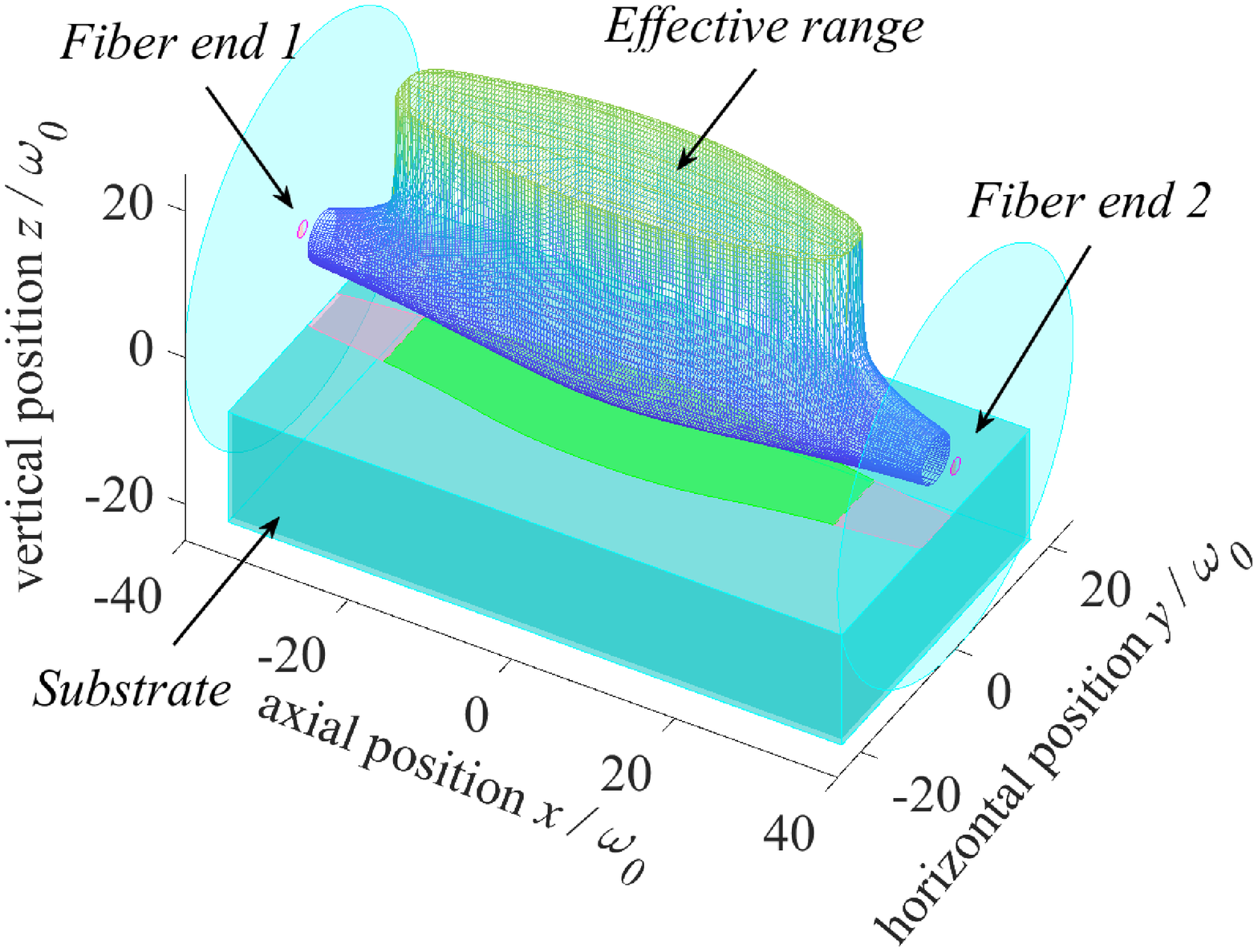}}\subfigure[]{\label{FIG3(c)}\includegraphics[width=0.47\textwidth]{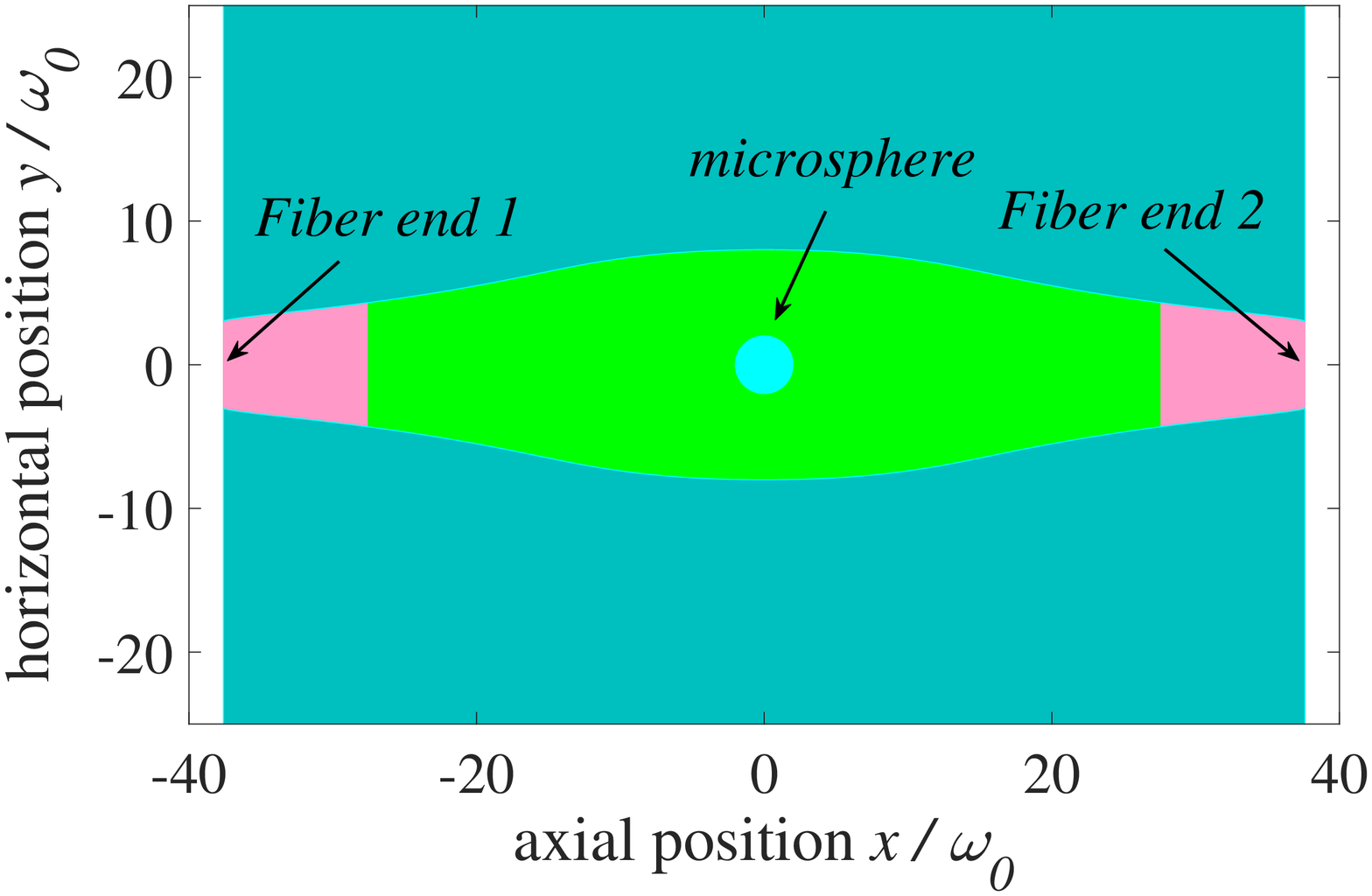}}
		\caption{(a) 3D vector graphical representation of the resultant forces. (b) 3D model displaying the effective capture region of the dual-fiber optical trap and the initial positions for a pulse laser launched microsphere. The effective region is covered by the mesh surface. A light blue round and a pink round surface on both sides respectively marks the fiber end and the beam waist according to the actual geometry. (c) The projection of effective region on the substrate. The pink and green part marks the projection, where a microsphere launched by pulse laser will enter the optical trap. And microspheres located in pink area are prone to stick to the end surface instead of being captured after launched by pulse laser. The green area is proposed as the initial loading positions of the microspheres.}
		\label{FIG4}
	\end{figure}
	
	The vector arrows in Fig.\ref{FIG3(a)} reveal the three-dimensional information about resultant forces. Based on the simulated force field, we obtain the effective capture region of the dual-fiber optical trap, as shown in Fig.\ref{FIG3(b)}. The effective capture region of dual-fiber optical trap is quite narrow compared to the space between the fibers, which makes it difficult for a launched microsphere to be captured. Therefore, the capture efficiency of traditional loading methods with shaking mechanism or nebulizer source are relatively low and uncontrollable because microspheres get into the space randomly. In addition, microspheres near the fiber ends are prone to strike the end surface because the effective region here is confined. It would block the exit of laser beam and further hinder the capture of other microspheres if a microsphere already adheres to the end. Consequently, to make sure the microsphere can be captured by CW lasers, its initial positions on the substrate should be limited to the green area shown in Fig.\ref{FIG3(c)} before it is launched by pulse laser.
	
	Similar to the cleaning threshold in the field of laser cleaning, the detachment threshold in this study is the minimum input laser energy required to launch the microspheres from the substrate. To reduce the detachment threshold and to make the capture process more efficient, some parameters of the pulse beam were analyzed, including the pulse durations, the size of beam waist, and waist position.
	\subsection{Suitable pulse duration for launch}
	According to Eq.\ref{eqFA}, the initial magnitude of adhesion forces is about $10^5$ times the magnitude of the gravity. Once the separation between the particle and substrate exceeds a threshold of $Z_{th}=59.8\,\rm{nm}$, the magnitude of adhesion forces will be smaller than gravity, and the microsphere will be free from the substrate easily. The duration of a pulse shot is so short that it cannot sustain the upward movement of microspheres all the time. The irradiation of pulse beam should not only provide a lifting force larger than the adhesion force in initial state, but also release the microspheres by a sufficient distance away from the substrate. Therefore, the minimum energy $E_m$ transferred from single pulse shot should be larger than the work done while lifting the microsphere against the adhesion forces: 
	\begin{equation}
	E_m\ge\int_{Z_0}^{Z_{th}}F_A\,dZ=\int_{Z_0}^{Z_{th}}{\frac{Aa}{6Z^2}}\,dZ
	\label{eq14}
	\end{equation}
	
\begin{figure}
		\centering
		\subfigure[]{\label{FIG4(a)}\includegraphics[width=0.5\textwidth]{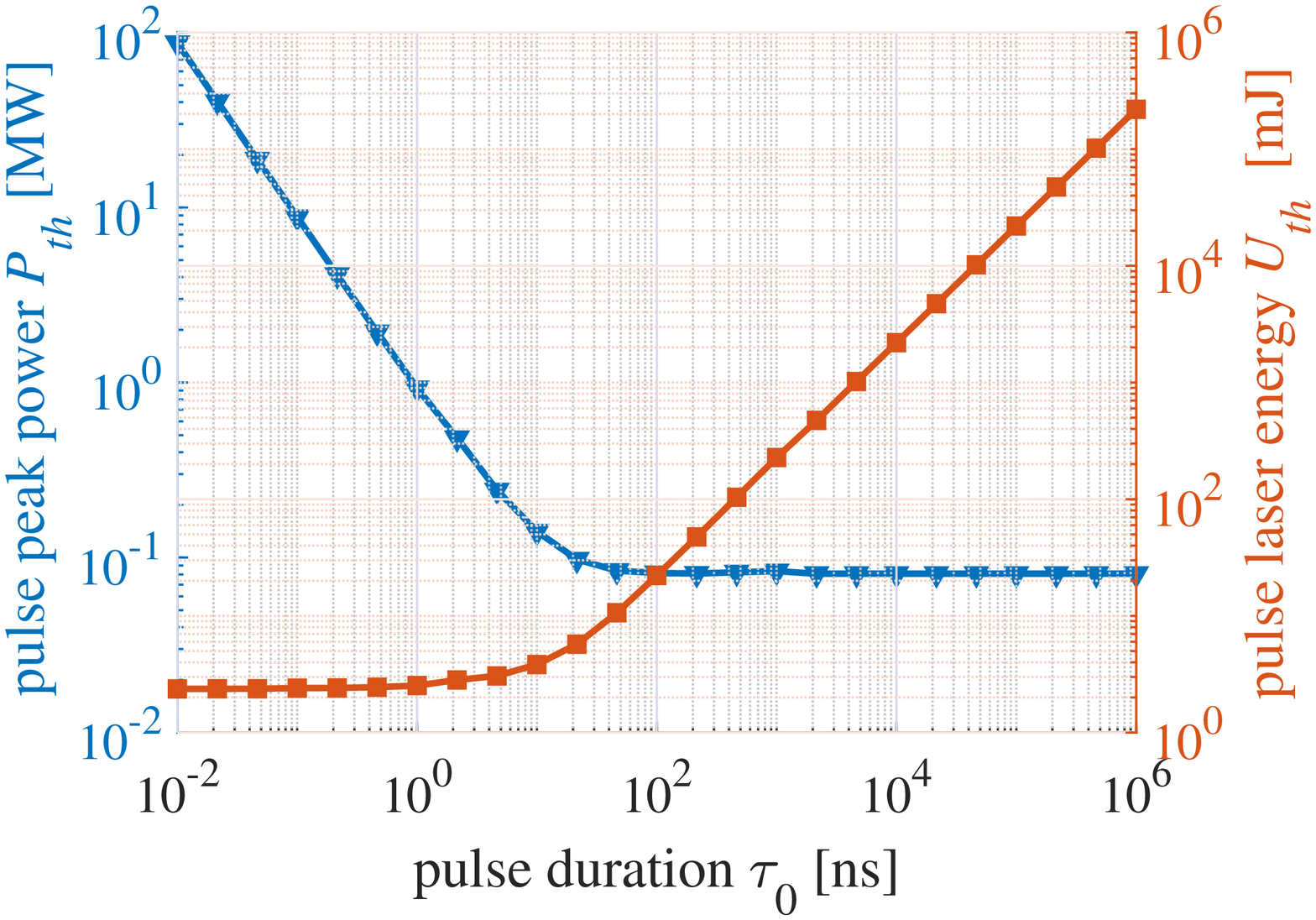}}\subfigure[]{\label{FIG4(b)}\includegraphics[width=0.5\textwidth]{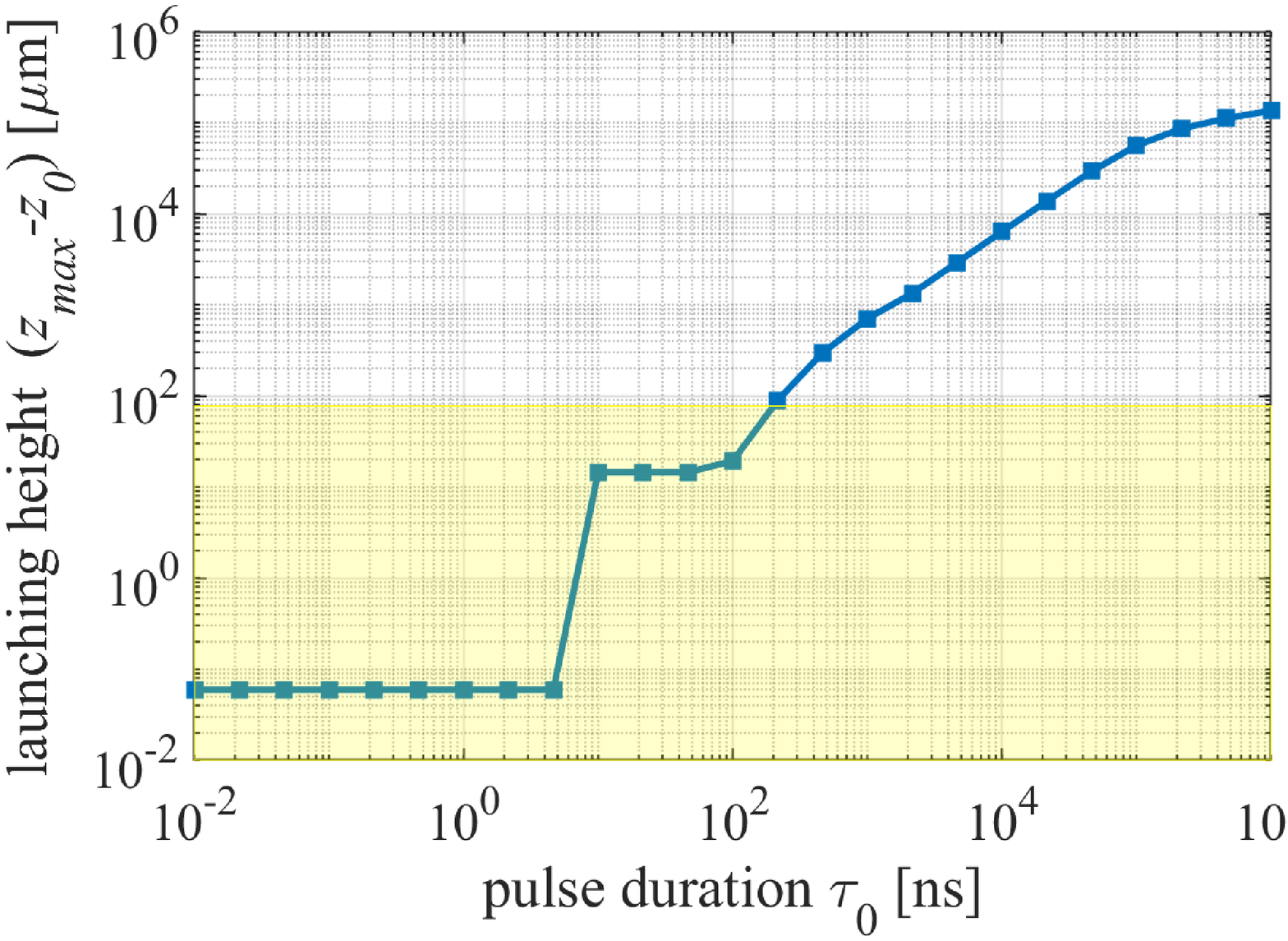}}
		\caption{(a) The calculated minimum required pulse peak power and pulse laser energy as functions of pulse duration (from 10 ps to 1 ms). The microsphere is initially located in the spot center, and the surface of the substrate have a vertical displacement $z_s=-20\,\rm{\upmu m}$ from the beam axis of optical trap. (b) The launching height as function of pulse duration. The launching height is defined as the distance between the highest position $z_{max}$ and the initial position $z_0$  after the microsphere is irradiated by the pulse beam as well as the CW beams. The yellow shaded area indicates the effective region of the CW trap.}
	\end{figure}

	In simulation, we choose $\omega_p=100\,\rm{\upmu m}$ as the waist of pulse beam so that its spot can cover the effective region shown in Fig.\ref{FIG3(c)}. The surface of the substrate is set to have a vertical displacement $z_s=-20\,\rm{\upmu m}$ from the beam axis of optical traps, thus microspheres can enter the effective region as soon as it is separated from the surface. Figure \ref{FIG4(a)} shows the minimum peak powers $P_{th}$ required for different pulse durations $\tau_0$ (from 10ps to 1ms) to extract a microsphere initially located in the spot center. And the detachment threshold of a single pulse shot can be obtained by the peak power and the pulse width as follow:
	\begin{equation}
	U_{th}=\int_{0}^{\infty}P_{th}(t)\,dt=\int_{0}^{\infty}P_{th}(t)\cdot\frac{t}{\tau_0}\cdot\mathrm{e}^{1-\frac{t}{\tau_0}} \,dt=\mathrm{e}\cdot\tau_0\cdot P_{th}
	\end{equation}
	
	The required minimum peak power decreases as the pulse width increases and gradually approaches a lower limit when the pulse duration is greater than 100 ns. If the peak power of the pulse light is lower than this limit, the radiation force will be insufficient to overcome the adhesive forces and gravity exerted on the microsphere. When the pulse duration is long enough, the amount of energy transmitted by the pulse beam is partially offset by the work indicated in Eq.\ref{eq14}, and the residual is converted into the initial kinetic energy of the microsphere. This could explain the fact that although the peak power is decreasing, the detachment threshold increases sharply with the pulse duration, which has been experimentally verified by Saki Maeda and Tadao Sugiura \cite{maeda2012optical}. 
	
	For a clearer understanding about the effect of pulse duration, we investigated the launching height as function of pulse duration in cases of corresponding detachment thresholds indicated in Fig.\ref{FIG4(a)}. As shown in Fig.\ref{FIG4(b)}, the launching height of the microspheres is beyond the upper boundary of the fiber if the pulse duration is longer than 110 ns. In those cases, the initial kinetic energy will be superfluous to the CW optical trap because microspheres can easily escape from the effective range of the optical trap. On the other hand, the pulse beam with pulse width shorter than 1 ns barely provides enough kinetic energy and the detachment threshold is relatively lower. But laser-induced dielectric breakdown phenomena will occur if the pulse duration is too short \cite{kane2006laser}. From this simulation result, we conclude that the suitable pulse duration is around 1 to 100 ns in the case of manipulation of micrometer-sized particles in a dual-beam fiber-optic trap under air condition.
	
	\subsection{The size of pulse beam waist}
\begin{figure}
		\centering
		\subfigure[]{\label{FIG5(a)}\includegraphics[width=0.5\textwidth]{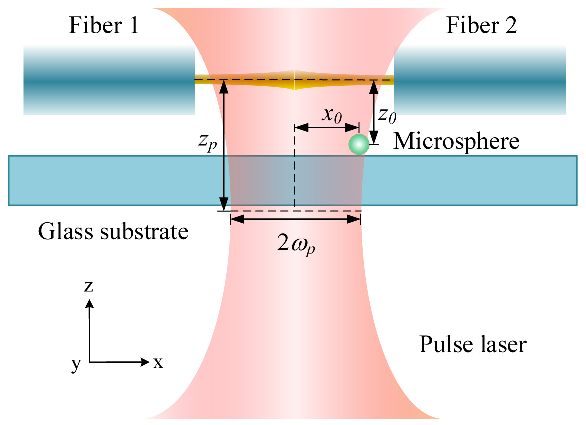}}\subfigure[]{\label{FIG5(b)}\includegraphics[width=0.5\textwidth]{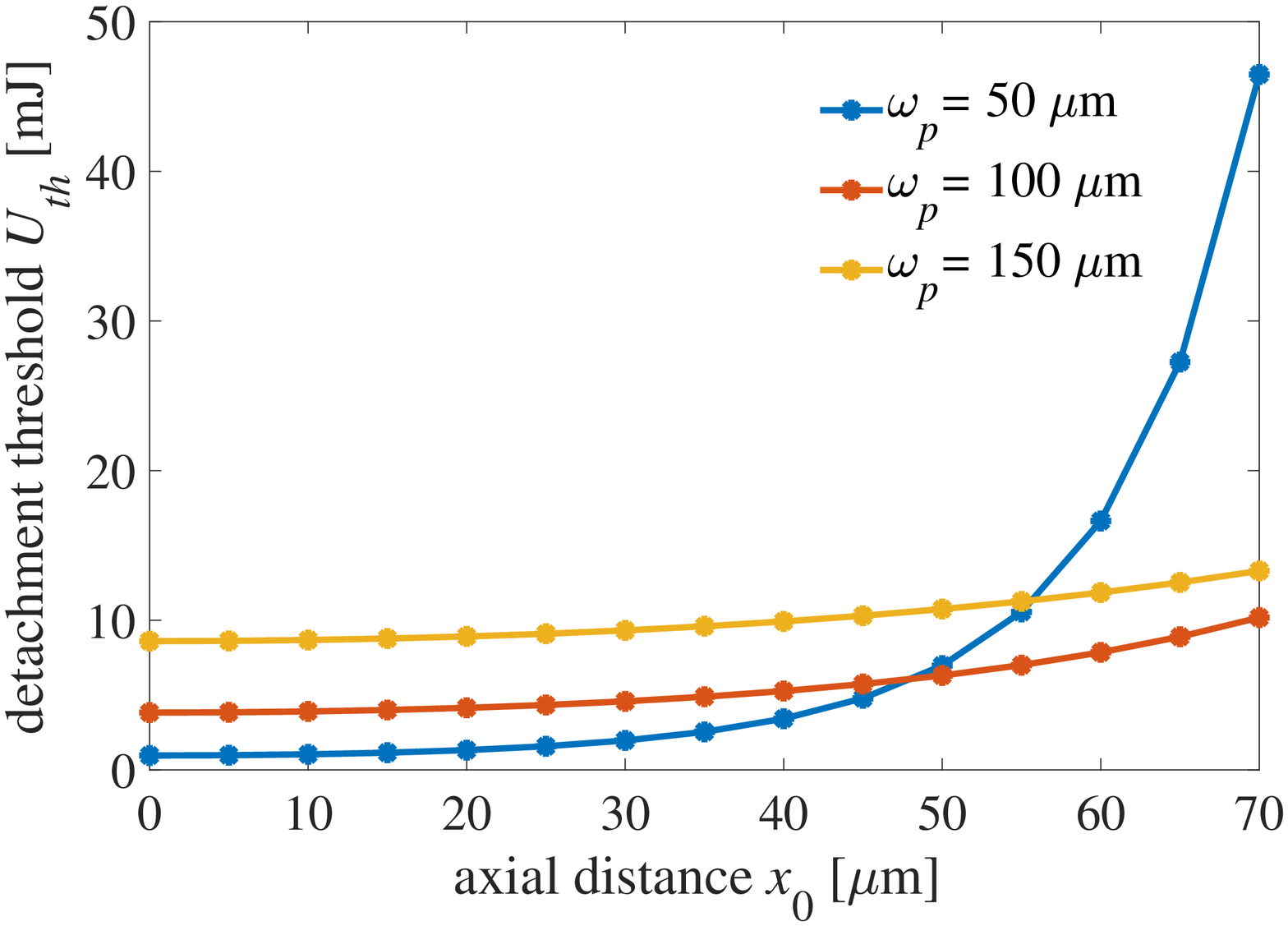}}
		\subfigure[]{\label{FIG5(c)}\includegraphics[width=0.5\textwidth]{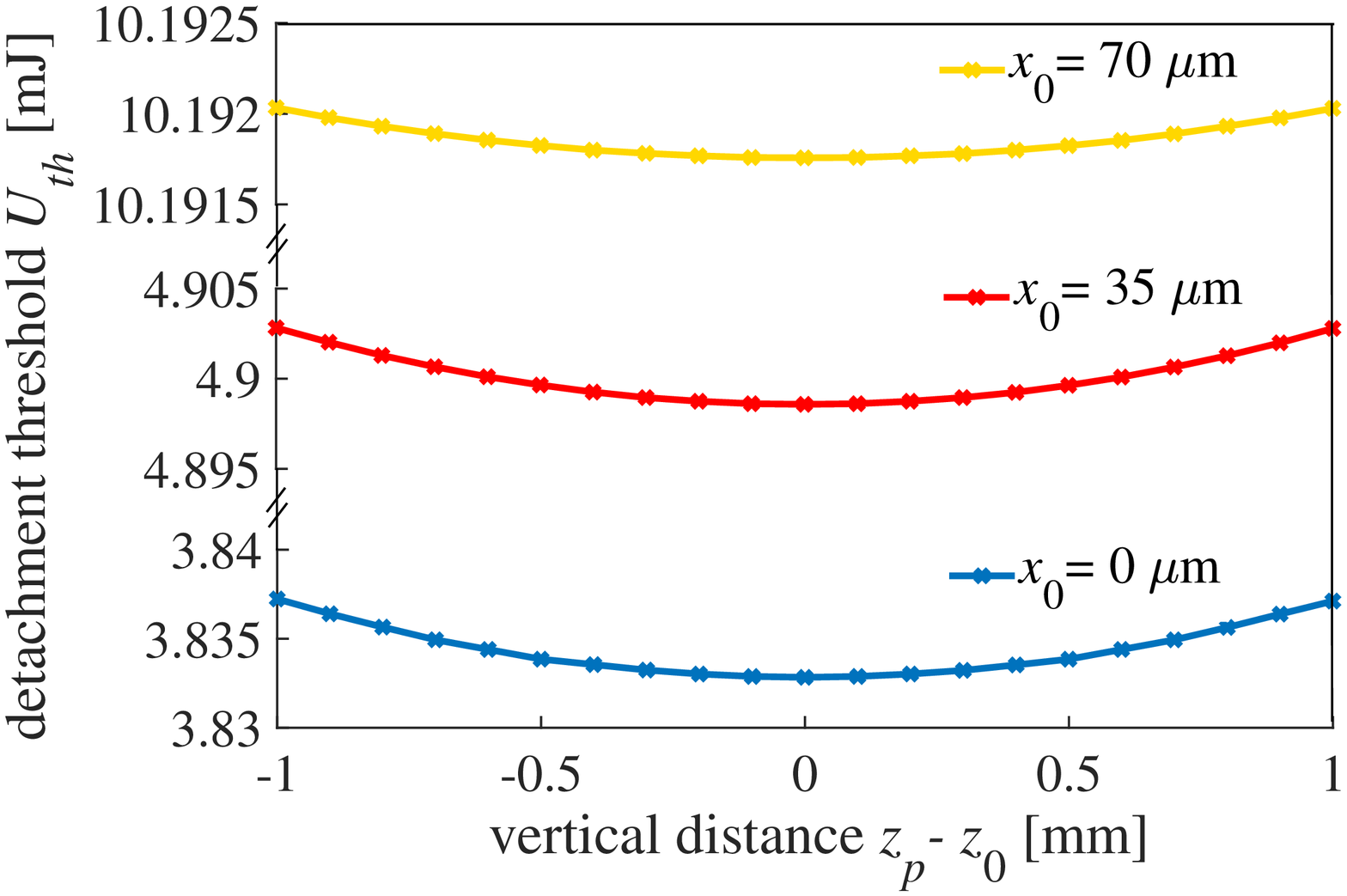}}\subfigure[]{\label{FIG5(d)}\includegraphics[width=0.5\textwidth]{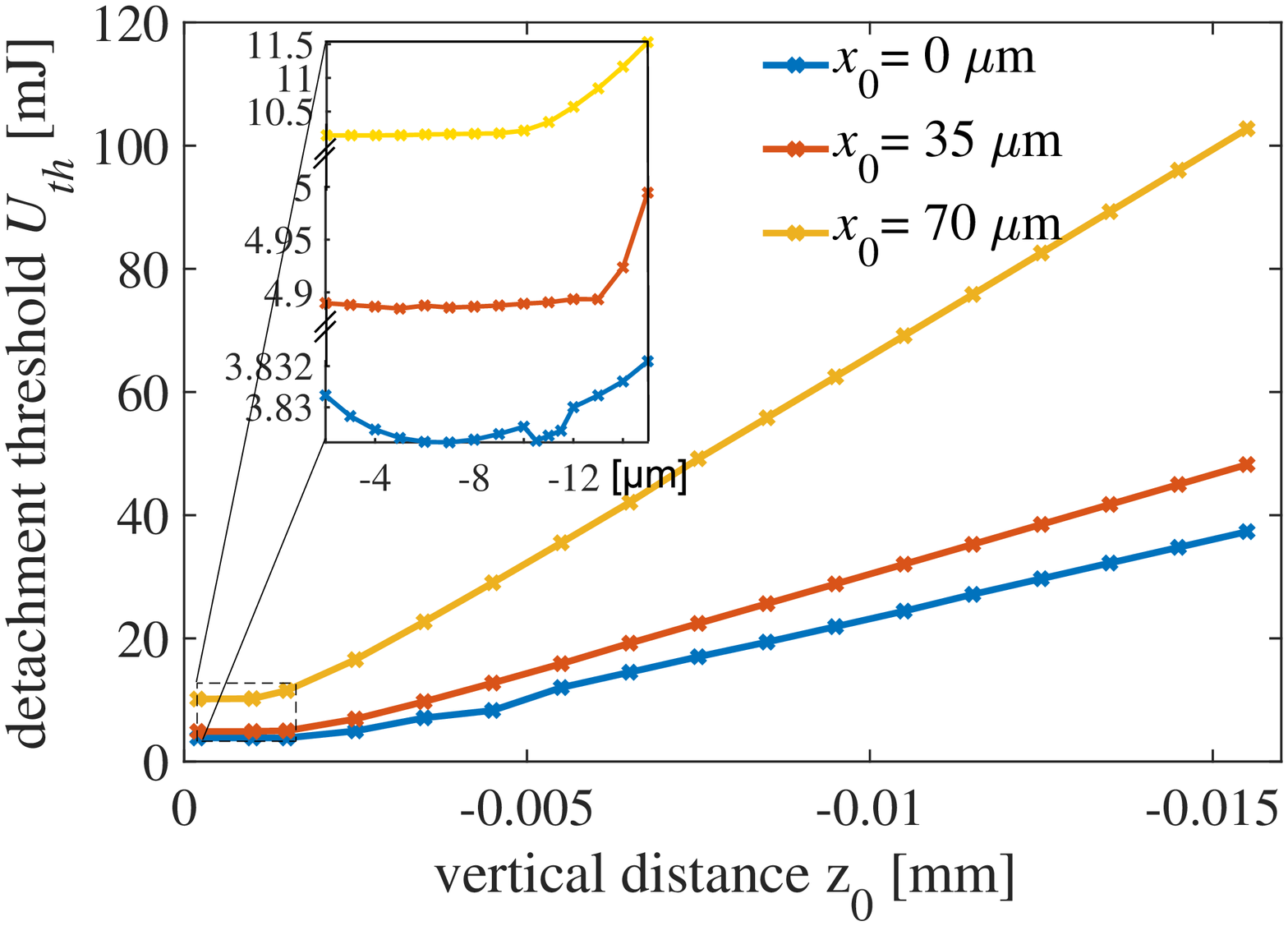}}
		\caption{(a) Schematic showing the structural parameters of PLAT. The detachment thresholds as function of different structural parameters: (b) the size of pulse beam waist $\omega_p$, (c) the vertical distance between pulse beam waist and microsphere $z_p-z_0$, (d) the vertical displacement between CW beam axis and microsphere $z_0$. The pulse duration is 10 ns in simulations. The pulse beam waist is 100\,$\rm{\upmu m}$ in (c) and (d). In (b) and (d), the beam waist and microsphere are in the same plane .}	
	\end{figure}

	To study the effect of beam spot size, we analyzed the launching process in a more complex 3D situation. Figure \ref{FIG5(b)} shows the detachment thresholds of different sizes of pulse beam waist with a fixed pulse duration $\tau_0=10\,\rm{ns}$. The detachment threshold increases with axial distance between initial position of microsphere and the spot center. However, the increase speeds up substantially as the axial distance exceeds beam waist. Obviously, for a smaller beam waist, the light energy will have a more concentrated distribution, and the detachment threshold will be lower for microspheres located near the spot center. But for those microspheres diverging from the spot center, the required detachment threshold will be much higher due to the rapid attenuation of the intensity. In experiment, the spot of pulse beam can be controlled by optical elements, and we can obtain a suitable size of waist according to practical situation.
	\subsection{The vertical displacement between pulse beam waist and microsphere}
	The microspheres are detached from the substrate primarily by the action of the scattering force component of pulse beam, which varies with different positions along it beam axis. In previous simulation, we assumed that the microsphere lies in the plane of the beam waist. However, it is quite difficult to guarantee this situation in experiment and there may exist a certain vertical distance between the microsphere and the pulsed beam waist. We studied the effect of their vertical distance on the detachment threshold. Figure \ref{FIG5(c)} shows the dependence of the detachment threshold on the vertical distance with different initial positions of microsphere. The detachment threshold reaches a minimum value when the vertical distance is zero, corresponding to the maximum scattering force generated by the pulse beam. Although the detachment threshold increases slightly when they don't overlap each other, the vertical distance has little impact on the detachment threshold. The simulation suggests that a precise alignment of the microsphere and the pulsed beam waist is dispensable in experiment, provided that their vertical distance does not exceed a significant value (e.g., 1mm).
	
	\subsection{The vertical displacement between CW beam axis and microsphere}
	The initial vertical displacement of the microsphere has a great effect on the detachment threshold. Figure \ref{FIG5(d)} shows the dependence of the detachment threshold on the vertical distance with different axial positions of microsphere. The microspheres initially within the effective region are irradiated by both CW beams and pulse beam, hence the detachment thresholds will be related to the distribution of gradient forces. On the other hand, for the microspheres initially outside the effective region, the detachment thresholds increase linearly with the initial vertical displacement of the microsphere because the pulse beam should provide sufficient kinetic energy for the microspheres to enter the effective region. Consequently, it can be proposed that microspheres are placed near the effective region for minimizing the pulse beam energy. 
	
	\section*{Conclusion}
	In this work, we have proposed a new structure of PLAT that combinates pulse laser with dual-beam fiber-optic trap. We quantitatively show the effective region of an aligned dual-beam fiber-optic trap, which offered a profound understanding of conditions for capturing a launched microsphere in free space. By analyzing the energy transfer process of pulse beam irradiation, we have derived a suitable pulse duration from 1 to 100 ns in the case of manipulation of micro-sized particles in a dual-beam fiber-optic trap in air. Parametric studies are conducted to get an optimal structure of PLAT with a fixed duration of 10 ns. We have found that an appropriate pulse beam waist size is necessary for efficient detachment, but the effect of the vertical displacement between waist and microsphere becomes negligible. The initial position of the microsphere is preferably near the effective capture region and directly below the equilibrium point of dual-beam fiber-optic trap, which reduces the pulse beam energy required for launch. Our numerical model and parametric studies are also applicable to other dual-beam optical traps of similar structure, such as free-space gradient force traps. Meanwhile, the replacement of other materials or sizes of microspheres does not radically affect our conclusions.
	
	Our result provides theoretical basis for the design of the PLAT, which will be a breakthrough for controllable microsphere manipulations in optical trap setup. By precisely controlling the movement state of a single launched microsphere, we hope to make the loading process more efficient, and ultimately to achieve a repeatable optical levitation.
	
	\section*{Funding Information}
	The Joint Fund of the Ministry of Education of China (6141A02011604); the National Natural Science Foundation of China (11304282, 10947104); the Fundamental Research Funds for the Central Universities (2016XZZX004-01).
	
	
	\bibliography{mybibfile}
	
\end{document}